**BIODIELEC06**

# Casimir interaction between gas media of excited atoms

**Yury Sherkunov**
Institute for High Energy Density of Joint Institute for High Temperatures, RAS, Moscow, Russia

sherkunovyb@physics.org

**Abstract**. The retarded dispersion interaction (Casimir interaction) between two dilute dielectric media at high temperatures is considered. The excited atoms are taken into account. It is shown that the perturbation technique can not be applied to this problem due to divergence of integrals. A non perturbative approach based on kinetic Green functions is implemented. We consider interaction between two atoms (one of them is excited) embedded in an absorbing dielectric medium. We take into account the possible absorption of photons in the medium, which solves the problem of divergence. The force between two plane dilute dielectric media is calculated at pair interaction approximation. We show that the result of quantum electrodynamics differs from the Lifshitz formula for dilute gas media at high temperatures (if the number of exited atoms is significant). According to quantum electrodynamics, the interaction may be either attractive or repulsive depending on the temperature and the density numbers of the media.

34.20.Cf, 12.20.-m, 42.50.Vk, 34.50.Dy

**I. Introduction.**
The dispersion forces – the electromagnetic forces between neutral unpolarized objects – has been widely studied since 1930, when Fritz London considered interaction between two ground-state atoms [1]. He attributed the attraction between two atoms to the fluctuations of their dipole moments. Later in 1948 Casimir and Polder considered the so-called retarded regime of interaction, where the finite velocity of light should be taken into account. They showed that the dispersion interaction for large distances between the atoms is due to vacuum fluctuations of electromagnetic field [2]. The dispersion interaction between two metal plates at zero temperature was considered by Casimir in 1948 [3]. Later these results were generalized to dispersion interaction between dielectric or metal macroscopic media at finite temperatures by E.M.Lifshitz and collaborates [4,5]. The interest to Casimir physics is still growing not only due to wide range of applications to various areas of physics, chemistry, and biology, but either due to numerous controversies in theoretical description of the effect. For reviews of recent progress and controversies in the Casimir physics see [6].
Dispersion forces play a significant role in understanding of many phenomena in biochemistry [7], these forces are responsible for climbing ability of geckos [8] and spiders [9].
In biophysics a significant role plays the dispersion interaction between excited and ground-state molecules and the related Foster resonance energy transfer (FRET) [10]. This interaction is responsible for energy transfer from chlorophyll molecules to photosynthesis centers [11]. FRET can serve as biophysical ruler. The rate of FRET indicates the distance between the molecules [12,13]. Singe-molecule studies are used to probe biomolecules and polymers [14]. The FRET process plays a significant role in energy exchange in proteins [15]. Some aspects of the excited systems out of equilibrium applied to FRET and dispersion forces are discussed in [16].
Recently dispersion interaction between two gas medium containing excited atoms was considered in [17]. The author studied only the non-retarded regime of the distances small in comparison with the wavelengths of atom transitions. It was shown that the Lifshitz formula for interaction between dilute gas media at temperatures high enough for the medium to contain excited atoms is in contradiction with experimental evidence [17, 18]. We should stress here that the Lifshitz formula was intended for macroscopic continuous media, but not for dilute gases. If the amount of excited atoms in the dilute medium is negligible, the results obtained with the help of Lifshitz formula coincide with the ones obtained using quantum electrodynamics. But if the amount of excited atoms is significant the results of two approaches do not coincide [17]. As it was shown in [17], the interaction between an excited atom and a dilute medium of ground-state atoms is resonant; the atom may either be attracted or repulsed by the medium depending on the frequencies of the excited atom and the atoms of the medium. The interaction may be up to several orders of magnitude more intense than the interaction between a ground-state atom and the medium. These properties of the interaction are in the qualitative agreement with the experimental evidence obtained by the French group [18] for the interaction between the Cs atoms and the sapphire wall. These results coincide with the theoretical predictions as well [19]

Here, we consider a general case (including both retarded and non-retarded regimes) of interaction between dilute gas media. We demonstrate the violation of the Lifshitz formula for any distances between the dilute gas media if the amount of excited atoms is significant.

In Section II we consider interaction between an excited atom and a ground-state one. We show that the perturbation method results in divergence of integrals for the interaction between an excited atom and an infinitely stretched cloud of ground-state atoms. We investigate interaction between an excited atom and a ground state one embedded in an absorbing dielectric medium using a non-perturbative approach developed in [17]. We take into account a possible absorption of photons in the medium. As a result, the interaction potential contains exponential factor depending on the imaginary part of the refractive index of the medium. Thus, the problem of divergence is solved.

In Section III we investigate interaction between two medium at high temperatures. We take into account thermal radiation. We show that if the amount of excited atoms in both media is significant, the result obtained with the help of quantum electrodynamics differs from the one obtained with the help of the Lifshitz formula for dilute gas medium at least for our model given by the Hamiltonian (1).

## II. Interaction between an excited atom and a ground-state one embedded in a dielectric absorbing medium

We consider interaction between two two-level atoms embedded in an absorbing dielectric medium. Let atom A be excited and atom B be in its ground state. First, we suppose that the thermal photons are absent.

The Hamiltonian of the system is as follows

$$\hat{H} = \hat{H}_A + \hat{H}_B + \hat{H}_{med} + \hat{H}_{ph} + \hat{H}_{int}, \qquad (1)$$

where $\hat{H}_A = \sum_i \varepsilon_{Ai} \hat{b}_i^\dagger \hat{b}_i$, $\hat{H}_B = \sum_i \varepsilon_{Bi} \hat{\beta}_i^\dagger \hat{\beta}_i$, $\hat{H}_{med} = \sum_i \varepsilon_{medi} \hat{c}_i^\dagger \hat{c}_i$ are the Hamiltonians of noninteracting atoms A, B, and the atoms of the medium, $\varepsilon_i$ is the energy of i-th state of the corresponding atom, $\hat{b}_i (\hat{b}_i^\dagger), \hat{\beta}_i (\hat{\beta}_i^\dagger), \hat{c}_i (\hat{c}_i^\dagger)$ are annihilation (creation) operators of i-th state of corresponding atom, $\hat{H}_{ph} = \sum_{k\lambda} \omega(\lambda) \left( \hat{\alpha}_{k\lambda}^\dagger \hat{\alpha}_{k\lambda} + \frac{1}{2} \right)$ is the Hamiltonian of free electromagnetic field, $k$ is the wave vector, $\lambda = 1, 2, 3$ is the index of polarization of electromagnetic field, $\hat{\alpha}_{k\lambda} \left( \hat{\alpha}_{k\lambda}^\dagger \right)$ are annihilation (creation) operators of electromagnetic field.

The interaction Hamiltonian in the interaction representation is

$$\hat{H}_{int\,l}(t) = -\int \hat{\psi}_l^\dagger(x) \hat{d}^\nu \hat{E}_l^\nu(x) \hat{\psi}_l(x) d\mathbf{r} - \int \hat{\varphi}_l^\dagger(x) \hat{d}^\nu \hat{E}_l^\nu(x) \hat{\varphi}_l(x) d\mathbf{r} - \int \hat{\chi}_l^\dagger(x) \hat{d}^\nu \hat{E}_l^\nu(x) \hat{\chi}_l(x) d\mathbf{r} \qquad (2)$$

where

$$\hat{\psi} = \sum_i \psi_i (\mathbf{r} - \mathbf{R}_A) e^{-i\varepsilon_{Ai} t} \hat{b}_i, \quad \hat{\varphi} = \sum_i \varphi_i (\mathbf{r} - \mathbf{R}_B) e^{-i\varepsilon_{Bi} t} \hat{\beta}_i, \quad \hat{\chi}_l(x) = \sum_i \chi_i (\mathbf{r} - \mathbf{R}_m) e^{-i\varepsilon_{medi} t} \hat{c}_i \qquad (3)$$

with $\psi_i (\mathbf{r} - \mathbf{R}_A)$, $\varphi_i (\mathbf{r} - \mathbf{R}_B)$, and $\chi_i (\mathbf{r} - \mathbf{R}_m)$ being the wave functions of i-th state of corresponding atoms. $\hat{d}^\nu$ is the operator of dipole moment, $\hat{E}^\nu(\mathbf{r})$ is the operator of free electromagnetic field

$$\hat{E}^\nu(x) = i \sum_{k\lambda} \sqrt{\frac{2\pi \omega(\lambda)}{V}} e_{k\lambda}^\nu \left( \hat{\alpha}_{k\lambda} e^{i\mathbf{k}\mathbf{r}} e^{-i\omega\lambda(\lambda)t} - \hat{\alpha}_{k\lambda}^\dagger e^{-i\mathbf{k}\mathbf{r}} e^{i\omega(\lambda)t} \right), \qquad (4)$$

where V is the quantization volume, $e_{k\lambda}^{\nu}$ is the polarization unit vector, $\omega(1,2)=k, \omega(3)=0$.
$\boldsymbol{R}_m$ describes the position of an atom of the dielectric medium, $\boldsymbol{R}_A$ and $\boldsymbol{R}_B$ are radius-vectors corresponding to the position of atom A and atom B, $x=\{\boldsymbol{r},t\}$.

We suppose that the lifetime of excited state of atom A is long in comparison with the one of atom B, thus, we can calculate the interaction potential of the atoms as the energy shift of, say, atom B due to the presence of atom A.

$$U(\boldsymbol{R}_A - \boldsymbol{R}_B) = \Delta E_B. \tag{5}$$

To calculate the potential, we will use the method of kinetic Green functions applied to quantum electrodynamics [17].
Let

$$G_{ll'}^{B}(x,x') = -i\left\langle \hat{T}_c \hat{\varphi}_l(x) \hat{\varphi}_{l'}^{\dagger}(x') \hat{S}_c \right\rangle$$

be the Green function of atom B.

$$\hat{S}_c = \hat{T}_c \exp\left\{ \sum_{l=1,2} (-1)^l i \int_{-\infty}^{\infty} \hat{H}_{int\,l}(t) dt \right\} \tag{6}$$

is the scattering operator introduced by Keldysh [20,21], $\langle ... \rangle$ means averaging over initial state of free atoms and vacuum state of the electromagnetic field. $\hat{T}_c$ is the operator of time-ordering [20,21]. It acts as follows

$$\hat{T}_c \hat{A}_1(t)\hat{B}_1(t') = \begin{cases} \hat{A}(t)\hat{B}(t'), & t>t' \\ \hat{B}(t')\hat{A}(t), & t<t' \end{cases}, \quad \hat{T}_c \hat{A}_2(t)\hat{B}_2(t') = \begin{cases} \hat{B}(t')\hat{A}(t), & t>t' \\ \hat{A}(t)\hat{B}(t'), & t<t' \end{cases},$$

$$\hat{T}_c \hat{A}_1(t)\hat{B}_2(t') = \hat{B}(t')\hat{A}(t), \quad \hat{T}_c \hat{A}_2(t)\hat{B}_1(t') = \hat{A}(t)\hat{B}(t').$$

The matrix of density of atom B is

$$\rho^B(x,x') = iG_{12}^B(x,x').$$

It can be represented [17] as a sum of contributions of two channels

$$\rho^B(x,x') = \rho_c^B(x,x') + \rho_n^B(x,x').$$

$\rho_c^B(x,x')$ represents the so-called coherent channel [22] (the atom does not change its initial state, e.g. elastic scattering), $\rho_n^B(x,x')$ represents the so-called incoherent channel [22] (the atom changes its initial state, e.g. spontaneous decay). For our purpose the incoherent channel plays no role, so we will skip this channel.

The matrix of density of the coherent channel obeys the following equation [17]

$$\rho_c^B(x,x'') = \Psi(x)\Psi^*(x''),$$

$$\left( i\frac{\partial}{\partial t} - \hat{H}_B \right)\Psi(x) = \int M_{11}(x,x')\Psi(x')dx',$$

Where $M_{11}(x,x')$ is the mass operator of the system.
In the pole approximation, the solution of the equation is [17]

$$\rho_c^B(E,E',\boldsymbol{r},\boldsymbol{r}') = \frac{\varphi_0(\boldsymbol{r}-\boldsymbol{R}_B)\varphi_0^*(\boldsymbol{r}'-\boldsymbol{R}_B)e^{i(E-E')t_0}}{\left(E-\varepsilon_{B0}-M_{11}^{00}(\varepsilon_{B0})\right)\left(E'-\varepsilon_{B0}-M_{22}^{00}(\varepsilon_{B0})\right)}.$$

Where $\varepsilon_{B0}$ is the energy of the initial state of atom B (including the Lamb shift), $\varphi_0(\boldsymbol{r}-\boldsymbol{R}_B)$ is the eigenfunctions of the initial state of atom B, $t_0$ is the time of switching on the interaction. The mass operator is

$$M_{11}^{00}(t,t_1) = \int \varphi_0^*(\boldsymbol{r}-\boldsymbol{R}_B) M_{11}(x,x_1) \varphi_0(\boldsymbol{r}_1-\boldsymbol{R}_B) d\boldsymbol{r} d\boldsymbol{r}_1 \tag{7}$$

And in the energy domain

$$M_{11}^{00}(\varepsilon_{B0}) = \int_{-\infty}^{\infty} M_{11}^{00}(t,t') e^{i\varepsilon_{B0}(t-t')} d(t-t').$$

Using equation (5) we can express the interaction potential via the real part of the mass operator

$$U(\mathbf{R}_A - \mathbf{R}_B) = \Delta E_B = \text{Re}\left[M_{11}^{00}(\varepsilon_{B0})\right].$$

The imaginary part of the mass operator corresponds to the collisional broadening of the excited level of the atom B.

To find the mass operator we need to expand the scattering matrix (6) and use the Wick theorem [23]. Taking into account that for a single atom the normal ordering

$$\left\langle \hat{N}\hat{\varphi}_{l_1}(x)\hat{\varphi}_{l_2}^\dagger(x')\hat{\varphi}_l^\dagger(x_1)\hat{\varphi}_{l'}(x_2)...\right\rangle = 0,$$

for any order but the second one, while the second order of normal product represents the density matrix of initial state of atom

$$\rho_0^B(x,x') = \left\langle \hat{N}\hat{\varphi}_{l_1}(x)\hat{\varphi}_{l_2}^\dagger(x')\right\rangle = \left\langle \hat{\varphi}^\dagger(x')\hat{\varphi}(x)\right\rangle.$$

As a result, for mass operators we arrive at the following expressions [17]

$$M_{11}(x,x') = -i\hat{d}^\nu\hat{d}^{\nu'} g_{11}^{0B}(x,x') D_{11}^{0\nu\nu'}(x',x),$$
$$M_{22}(x,x') = i\hat{d}^\nu\hat{d}^{\nu'} g_{22}^{0B}(x,x') D_{22}^{0\nu\nu'}(x',x). \tag{8}$$

The vacuum propagators of atom B are [21]

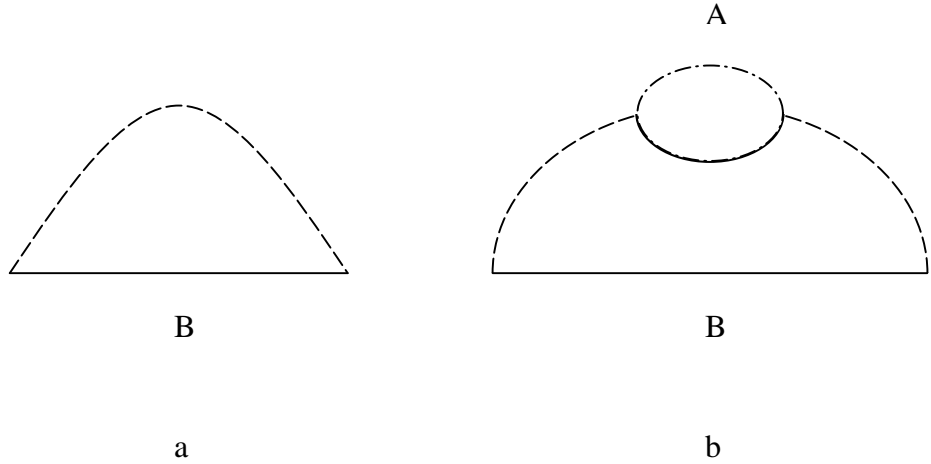

Fig.1. Feynman's diagrams for the mass operator (20). Solid line corresponds to atom propagator $g^0$. Dashed line corresponds to photon propagator $D^0$. Dashed-dotted line represents density matrix $\rho^0$.

$$g_{11}^{0B}(x,x') = -i\left\langle \hat{T}_c\hat{\varphi}_1(x)\hat{\varphi}_1^\dagger(x')\right\rangle_{vac} = -i\theta(t-t')\sum_i \varphi_i^*(\mathbf{r}')\varphi_i(\mathbf{r}) e^{-i\varepsilon_{Bi}(t-t')},$$

$$g_{22}^{0B}(x,x') = -i\left\langle \hat{T}_c\hat{\varphi}_2(x)\hat{\varphi}_2^\dagger(x')\right\rangle_{vac} = -i\theta(t'-t)\sum_i \varphi_i^*(\mathbf{r}')\varphi_i(\mathbf{r}) e^{-i\varepsilon_{Bi}(t-t')},$$

$$g_{12}^{0B}(x,x') = -i\left\langle \hat{\varphi}_2^\dagger(x')\hat{\varphi}_1(x)\right\rangle_{vac} = 0,$$

$$g_{21}^{0B}(x,x') = -i\left\langle \hat{\varphi}_2(x)\hat{\varphi}_1^\dagger(x')\right\rangle_{vac} = -i\sum_i \varphi_i^*(\mathbf{r}')\varphi_i(\mathbf{r}) e^{-i\varepsilon_{Bi}(t-t')}. \tag{9}$$

The expressions for the Green tensors of the photons read [21]

$$D_{ll'}^{0vv'}(x',x) = i\langle \hat{T}_c \hat{E}_l^v(x') \hat{E}_{l'}^{v'}(x) \rangle_{vac}. \qquad (10)$$

The Feynman diagram corresponding to the mass operators (8) are given in Fig.1.a This simplest mass operator describes the Lamb shift of atom B due to the interaction with the vacuum.

Taking into account the higher orders of perturbation technique and summing up an infinite sequence of the Feynman diagrams we find [17]

$$\begin{aligned} M_{11}(x,x') &= -i\hat{d}^v \hat{d}^{v'} g_{11}^B(x,x') D_{11}^{vv'}(x',x), \\ M_{22}(x,x') &= i\hat{d}^v \hat{d}^{v'} g_{22}^B(x,x') D_{22}^{vv'}(x',x). \end{aligned} \qquad (11)$$

Now $g_{11}^B(x,x')$ and $D_{11}^{vv'}(x',x)$ are the complete Green functions, which obey the Dyson equation [17,23]

$$\begin{aligned} g_{11}^B(x,x') &= g_{11}^{0B}(x,x') + \int dx_1 dx_2 g_{11}^{0B}(x,x_1) M_{11}(x_1,x_2) g_{11}^B(x_2,x'), \\ g_{22}^B(x,x') &= g_{22}^{0B}(x,x') + \int dx_1 dx_2 g_{22}^{0B}(x,x_1) M_{22}(x_1,x_2) g_{22}^B(x_2,x'). \end{aligned} \qquad (12)$$

The solution of this equation with the mass operator (11) is

$$g_{11}^B(E,\mathbf{r},\mathbf{r}') = \sum_i \frac{\varphi_i(\mathbf{r}-\mathbf{R}_B)\varphi_i^*(\mathbf{r}'-\mathbf{R}_B)}{E - \varepsilon_{Bi} + i\frac{\gamma_{Bi}}{2}}.$$

Where $\gamma_{Bi}$ is the width of state i of atom B.

$$g_{11}^A(E,\mathbf{r},\mathbf{r}') = \sum_i \frac{\psi_i(\mathbf{r}-\mathbf{R}_A)\psi_i^*(\mathbf{r}'-\mathbf{R}_A)}{E - \varepsilon_{Ai} + i0}.$$

$$\begin{aligned} D_{11}^{vv'}(x',x) &= D_{11}^{0vv'}(x',x) + \int dx_1 dx_2 \sum_{v_1 v_2} D_{11}^{0vv_1}(x',x_1) \Pi_{11}^{v_1 v_2}(x_1,x_2) D_{11}^{v_2 v'}(x_2,x) \\ &+ \int dx_1 dx_2 \sum_{v_1 v_2} D_{11}^{0vv_1}(x',x_1) \Pi_{12}^{v_1 v_2}(x_1,x_2) D_{21}^{v_2 v'}(x_2,x) \\ &+ \int dx_1 dx_2 \sum_{v_1 v_2} D_{12}^{0vv_1}(x',x_1) \Pi_{21}^{v_1 v_2}(x_1,x_2) D_{11}^{v_2 v'}(x_2,x), \end{aligned} \qquad (13)$$

$$D_{22}^{vv'}(x',x) = -D_{11}^{*v'v}(x,x').$$

Where $\Pi$ is the polarization operator.

$$\begin{aligned} \Pi_{11}^{v_1 v_2}(x_1,x_2) &= -\hat{d}^{v_1} \hat{d}^{v_2}\left(\rho^A(x_1,x_2) g_{11}^A(x_2,x_1) + g_{11}^A(x_1,x_2)\rho^A(x_2,x_1)\right) + \Pi_{11med}^{v_1 v_2}(x_1,x_2), \\ \Pi_{12}^{v_1 v_2}(x_1,x_2) &= \hat{d}^{v_1} \hat{d}^{v_2}\left(\rho^A(x_1,x_2) g_{21}^A(x_2,x_1) + g_{12}^A(x_1,x_2)\rho^A(x_2,x_1)\right) + \Pi_{12med}^{v_1 v_2}(x_1,x_2), \\ \Pi_{21}^{v_1 v_2}(x_1,x_2) &= \hat{d}^{v_1} \hat{d}^{v_2}\left(\rho^A(x_1,x_2) g_{12}^A(x_2,x_1) + g_{21}^A(x_1,x_2)\rho^A(x_2,x_1)\right) + \Pi_{21med}^{v_1 v_2}(x_1,x_2). \end{aligned} \qquad (14)$$

$\Pi_{med}^{v_1 v_2}$ is the polarization operator of the medium

The equation (13) can be rewritten in terms of the Green tensor of photons in the medium $D_{med}^{vv'}$

$$\begin{aligned} D_{11}^{vv'}(x',x) &= D_{11med}^{vv'}(x',x) + \int dx_1 dx_2 \sum_{v_1 v_2} D_{11med}^{vv_1}(x',x_1) \tilde{\Pi}_{11}^{v_1 v_2}(x_1,x_2) D_{11med}^{v_2 v'}(x_2,x) \\ &+ \int dx_1 dx_2 \sum_{v_1 v_2} D_{11med}^{vv_1}(x',x_1) \tilde{\Pi}_{12}^{v_1 v_2}(x_1,x_2) D_{21med}^{v_2 v'}(x_2,x) \\ &+ \int dx_1 dx_2 \sum_{v_1 v_2} D_{12med}^{vv_1}(x',x_1) \tilde{\Pi}_{21}^{v_1 v_2}(x_1,x_2) D_{11med}^{v_2 v'}(x_2,x). \end{aligned} \qquad (15)$$

$\tilde{\Pi} = \Pi - \Pi_{med}$ is the polarization operator of atom A without the medium.

It is convenient to introduce the retarded and advanced Green functions [21,23]

$$g_r = g_{11} - g_{12} = g_{21} - g_{22},$$
$$g_a = g_{11} - g_{21} = g_{12} - g_{22}.$$
$$D_r = D_{11} - D_{12} = D_{21} - D_{22},$$
$$D_a = D_{11} - D_{21} = D_{12} - D_{22}.$$
$$\Pi_r = \Pi_{11} + \Pi_{12} = \Pi_{22} + \Pi_{21},$$
$$\Pi_a = \Pi_{11} - \Pi_{21} = \Pi_{22} - \Pi_{12}.$$

(16)

The retarded Green functions have no poles in upper complex plane of energy, while the advanced Green functions are analytic in lower plane.

The vacuum Green tensor of the photons obeys the Dyson equation [23]

$$D_{r\,med}^{vv'}(x',x) = D_r^{0vv'}(x',x) + \int dx_1 dx_2 \sum_{v_1 v_2} D_r^{0vv_1}(x',x_1) \Pi_{r\,med}^{v_1 v_2}(x_1,x_2) D_{r\,med}^{v_2 v'}(x_2,x). \quad (17)$$

Since the polarization of the medium can be expressed in terms of the permittivity of the medium [24]

$$\Pi_{r\,med}^{vv'}(k,\omega) = \int \exp\left(-i k(r-r') + i\omega(t-t')\right) \Pi_{r\,med}^{vv'}(r-r',t-t') d(r-r') d(t-t')$$
$$= \frac{\delta_{vv'}(\varepsilon(\omega)-1)\omega^2}{4\pi},$$

we can easily write the solution of the equation (17) for infinite homogeneous medium [24]

$$D_{r\,med}^{vv'}(k,\omega) = -4\pi\omega^2 \left(\delta_{vv'} - \frac{k_v k_{v'}}{\varepsilon(\omega)\omega^2}\right)\left(\varepsilon(\omega)\omega^2 - k^2\right)^{-1}.$$

In the frequency-coordinate domain

$$D_{r\,med}^{vv'}(\omega,r-r') = \frac{1}{(2\pi)^3} \int_{-\infty}^{\infty} D_{r\,med}^{vv'}(\omega,k) \exp(ik(r-r')) d(r-r')$$

$$= \omega^2 \left[\delta_{vv'}\left(1 + \frac{i}{n(\omega)\omega|r-r'|} - \frac{1}{n^2(\omega)\omega^2|r-r'|^2}\right)\right.$$

$$\left. + \frac{(r-r')_v (r-r')_{v'}}{|r-r'|^2}\left(\frac{3}{n^2(\omega)\omega^2|r-r'|^2} - \frac{3i}{n(\omega)\omega|r-r'|} - 1\right)\right] \frac{e^{in(\omega)\omega|r-r'|}}{|r-r'|},$$

$$D_a^{vv'}(\omega,r-r') = \left(D_r^{vv'}(\omega,r-r')\right)^*$$

(18)

where

$$n(\omega) = \sqrt{\varepsilon(\omega)} \quad (19)$$

is the complex refractive index of the medium.

Substituting (15) and (14) into (11) and taking into account (7), we find

$$M_{11}^{00}(t,t') = -i\int \varphi_0^*(r-R_B)\hat{d}^v \hat{d}^{v'} g_{11}^{0B}(x,x') D_{11}^{0vv'}(x',x) \varphi_0(r_1-R_B) dr dr' +$$

$$-i\int \varphi_0^*(r-R_B)\hat{d}^v \hat{d}^{v'} g_{11}^B(x,x') \int dx_1 dx_2 D_{11\,med}^{vv_1}(x',x_1) \tilde{\Pi}_{11}^{v_1 v_2}(x_1,x_2) D_{11\,med}^{v_2 v'}(x_2,x) \varphi_0(r'-R_B) dr dr'$$

$$-i\int \varphi_0^*(r-R_B)\hat{d}^v \hat{d}^{v'} g_{11}^B(x,x') \int dx_1 dx_2 D_{11\,med}^{vv_1}(x',x_1) \tilde{\Pi}_{12}^{v_1 v_2}(x_1,x_2) D_{21\,med}^{v_2 v'}(x_2,x) \varphi_0(r'-R_B) dr dr'$$

$$-i\int \varphi_0^*(r-R_B)\hat{d}^v \hat{d}^{v'} g_{11}^B(x,x') \int dx_1 dx_2 D_{12\,med}^{vv_1}(x',x_1) \tilde{\Pi}_{21}^{v_1 v_2}(x_1,x_2) D_{11\,med}^{v_2 v'}(x_2,x) \varphi_0(r'-R_B) dr dr'.$$

(20)

The Feynman diagrams corresponding to the mass operator (20) are shown in Fig.1
The diagram in Fig.1(a) describes the interaction of atom B with the electromagnetic vacuum. It corresponds to the Lamb shift of the levels of atom B. The diagram in Fig.1(b) describes the energy shift of atom B due to the presence of atom A (dispersion interaction). The processes are as follows. The excited atom A emits a photon, which is absorbed by atom B, then atom B emits the photon, which, in its turn, is absorbed by the atom A. The final states of atoms do not change. After simple but somewhat lengthy calculations, we find

$$M_{11}^{00} =$$
$$\frac{i}{2\pi}\left[\frac{1}{2}\int_{-\infty}^{\infty}\alpha_{eA}^{(c)v_1v_2}(\omega)\alpha_{gB}^{(c)vv'}(\omega)D_{11med}^{vv_1}(\omega,\mathbf{R}_A-\mathbf{R}_B)D_{11med}^{v_2v'}(\omega,\mathbf{R}_A-\mathbf{R}_B)d\omega \right. \tag{21}$$
$$\left. -2\pi i\int_0^{\infty}d_{eg}^{Av_1}d_{eg}^{Av_2}\alpha_{gB}^{(c)vv'}\delta(\omega_A-\omega)D_{11med}^{vv_1}(\omega,\mathbf{R}_A-\mathbf{R}_B)D_{21med}^{v_2v'}(\omega,\mathbf{R}_A-\mathbf{R}_B)\theta(\omega_A)d\omega\right].$$

$\omega_A$ and $\omega_B$ are the transition frequencies of atoms A and B.

$\theta(\omega_A)$ is the unit step-function. If atom A is excited $\theta(\omega_A)=1$, if atom A is in its ground state $\theta(-\omega_A)=0$.

$\alpha^{(c)vv'}$ is the so-called coherent polarizability of an atom [23], which describes the coherent scattering of light on the atom.

For the excited atom, the coherent polarizability is

$$\alpha_e^{(c)vv'}(\omega)=\frac{d_{eg}^v d_{ge}^{v'}}{-\omega_{eg}-\omega-i\frac{\gamma}{2}}+\frac{d_{ge}^v d_{eg}^{v'}}{-\omega_{eg}+\omega-i\frac{\gamma}{2}}.$$

For the ground-state atom

$$\alpha_g^{(c)vv'}(\omega)=\frac{d_{ge}^v d_{eg}^{v'}}{\omega_{eg}-\omega-i\frac{\gamma}{2}}+\frac{d_{eg}^v d_{ge}^{v'}}{\omega_{eg}+\omega-i\frac{\gamma}{2}}.$$

Where $\omega_{eg}$ is the transition frequency of the atom.

Formula (21) can be rewritten using the conventional polarizabilities of the atoms

$$U(\mathbf{R}_A-\mathbf{R}_B)=\operatorname{Re}M_{11}^{00}$$
$$\operatorname{Re}\frac{i}{2\pi}\left[\int_0^{\infty}\alpha_{eA}^{v_1v_2}(\omega)\alpha_{gB}^{(c)vv'}(\omega)D_{11med}^{vv_1}(\omega,\mathbf{R}_A-\mathbf{R}_B)D_{11med}^{v_2v'}(\omega,\mathbf{R}_A-\mathbf{R}_B)d\omega\right.$$
$$\left.+2\pi i\int_0^{\infty}d_{eg}^{Av}d_{ge}^{Av'}\alpha_{gB}^{(c)vv'}(\omega)\delta(\omega_A-\omega)D_{11med}^{vv_1}(\omega,\mathbf{R}_A-\mathbf{R}_B)\left(D_{11med}^{v_2v'}(\omega,\mathbf{R}_A-\mathbf{R}_B)-D_{21med}^{v_2v'}(\omega,\mathbf{R}_A-\mathbf{R}_B)\right)\theta(\omega_A)d\omega\right].$$

Where the conventional polarizabilities are [23]

$$\alpha_g^{vv'}(\omega)=\frac{d_{ge}^v d_{eg}^{v'}}{\omega_{eg}-\omega-i\frac{\gamma}{2}}+\frac{d_{eg}^v d_{ge}^{v'}}{\omega_{eg}+\omega+i\frac{\gamma}{2}}, \tag{22}$$

$$\alpha_e^{vv'}(\omega)=\frac{d_{eg}^v d_{ge}^{v'}}{-\omega_{eg}-\omega-i\frac{\gamma}{2}}+\frac{d_{ge}^v d_{eg}^{v'}}{-\omega_{eg}+\omega+i\frac{\gamma}{2}}. \tag{23}$$

After averaging over all possible mutual orientations of dipole moments of atoms, we can write [23]

$$d_{eg}^{v_1}d_{ge}^{v_2}\to\frac{|d_{eg}|^2}{3}\delta_{v_1v_2} \tag{24}$$

Taking into account the symmetry properties of the Green tensors $D_a(\omega)=D_r^*(\omega)$ and $D_{12}=0$ for $\omega>0$, which could be checked directly using (10), and using (16), we find

$$U(\mathbf{R}_A - \mathbf{R}_B) = \mathrm{Re}\frac{\mathrm{i}}{2\pi}\left[\int_0^\infty \alpha_{eA}(\omega)\alpha_{gB}(\omega)\left(D_{r\,med}^{\nu\nu'}(\omega,\mathbf{R}_A-\mathbf{R}_B)\right)^2 d\omega \right.$$
$$\left. +\frac{2\pi\mathrm{i}}{3}\left|d_{eg}^A\right|^2 \alpha_{gB}(\omega_A)\left|D_{r\,med}^{\nu\nu'}(\omega_A,\mathbf{R}_A-\mathbf{R}_B)\right|^2 \theta(\omega_A).\right] \quad (25)$$

Finally, using (18) we have

$$U(R) = \mathrm{Re}\frac{\mathrm{i}}{\pi}\left[\int_0^\infty \alpha_{eA}(\omega)\alpha_{gB}(\omega)\frac{\omega^4}{R^2}\left(1+\frac{2\mathrm{i}}{n(\omega)\omega R}-\frac{5}{(n(\omega)\omega R)^2}\right.\right.$$
$$\left.-\frac{6\mathrm{i}}{(n(\omega)\omega R)^3}+\frac{3}{(n(\omega)\omega R)^4}\right)\exp(2\mathrm{i}n(\omega)\omega R)d\omega \quad (26)$$
$$\left.+\frac{2\pi\mathrm{i}}{3}\left|d_{eg}^A\right|^2\alpha_{gB}(\omega_A)\frac{\omega_A^4}{R^2}\left(1+\frac{1}{(n(\omega_A)\omega_A R)^2}+\frac{3}{(n(\omega_A)\omega_A R)^4}\right)\exp\left[-2\mathrm{Im}(n(\omega_A))\omega_A R\right]\theta(\omega_A)\right].$$

We should mention here, that for excited atom the interaction potential can not be expressed only in terms of polarizabilities of the medium. The second term of (26) contains only the polarizability of the ground-state atom, but not of the excited one. The exponent in the second term of (26) is due to possible absorption of photons by the medium.

Let us suppose that atoms A and B are in a vacuum. Then we should put $n(\omega)=1$. If we neglect the width of excited level of atom B, we will come to the result obtained by E. A. Power and T. Thirunamachandran using perturbation technique [25,26].
For the non-retarded limit ($R \ll \lambda$), where $\lambda$ is the wavelength of atom transition,

$$U(R) = -\frac{4}{3}\frac{\left|d_{eg}^A\right|^2\left|d_{eg}^B\right|^2 \omega_B}{(\omega_B^2-\omega_A^2)R^6}\theta(\omega_A). \quad (27)$$

For the retarded limit ($R \gg \lambda$), the result is

$$U(R) = -\frac{4}{9}\frac{\left|d_{eg}^A\right|^2\left|d_{eg}^B\right|^2 \omega_B \omega_A^4}{(\omega_B^2-\omega_A^2)R^2}\theta(\omega_A). \quad (28)$$

The most significant features of the formulas (27) and (28) are as follows. As it was shown in [25, 26], the interaction between and excited atom and a ground-state one is resonant. It may be either attractive, or repulsive depending on the frequencies of the atoms. For the retarded limit ($R \gg \lambda$) the interaction potential drops as $R^{-2}$ with the distance between the atoms, while for two ground-state atoms the dependence of the potential on the distance is $U \Box R^{-7}$ [2].

Let us consider interaction between an excited atom and a dilute gas cloud of ground-state atoms. Let the interface be a plane. The cloud stretches to infinity. For dilute gas we can take into account only pair interactions between the atoms. If we take the expression (28) obtained with the help of perturbation technique, we will find

$$U = -\frac{4}{9}\frac{\left|d_{eg}^A\right|^2\left|d_{eg}^B\right|^2 \omega_A^4}{(\omega_B^2-\omega_A^2)}n_0\int\frac{dV}{R^2} \to \infty, \quad (29)$$

where $n_0$ is the density number of the medium.
Thus, we come to the divergence. It means that one should use a non-perturbative approach to calculate the force between an excited atom and a dilute medium of ground-state atoms. Namely, we should substitute expression (26) into (29) but not the formula (28).
Let us consider a case of interaction between two ground-state atoms.

We should substitute $\omega_A \to -\omega_A$ into (26). The second term disappears.

$$U(R) = \text{Re}\frac{i}{\pi}\left[\int_0^\infty \alpha_{eA}(\omega)\alpha_{gB}(\omega)\frac{\omega^4}{R^2}\left(1 + \frac{2i}{n(\omega)\omega R} - \frac{5}{(n(\omega)\omega R)^2}\right.\right.$$
$$\left.\left. - \frac{6i}{(n(\omega)\omega R)^3} + \frac{3}{(n(\omega)\omega R)^4}\right)\exp(2in(\omega)\omega R)d\omega\right].$$

This formula can be rewritten using contour integration

$$U(R) = \text{Re}\left[-\frac{1}{\pi}\int_0^\infty \alpha_{gA}(iu)\alpha_{gB}(iu)\frac{u^4}{R^2}\right.$$
$$\left.\times\left(1 + \frac{2}{n(iu)uR} + \frac{5}{(n(iu)uR)^2} + \frac{6}{(n(iu)uR)^3} + \frac{3}{(n(iu)uR)^4}\right)\exp(-2n(iu)uR)du\right].$$

For two limiting cases we find

$$U(R) = -\text{Re}\left[\frac{3}{\pi}\int_0^\infty \frac{\alpha_{gA}(iu)\alpha_{gB}(iu)}{R^6(n(iu))^4}du\right], \quad R \ll \lambda, \qquad (30)$$

$$U(R) = -\frac{23\alpha_{gA}(0)\alpha_{gB}(0)}{4\pi(n(0))^5 R^7}, \quad R \gg \lambda. \qquad (31)$$

The expression (30) and (31) coincide with the corresponding formulae, obtained for two ground-state atoms embedded in a dielectric [27].

Now, let us take into account the thermal radiation in the medium. Such a problem has been solved by several authors for a case of two ground-state atoms without dielectric medium [28-33].

To obtain the interaction potential at finite temperatures G. Goedecke and R. Wood made a substitution

$$1/2 \to (N_{k\lambda} + 1/2) \qquad (32)$$

into the potential at zero temperatures [30], where $N_{k\lambda}$ is the number of photons in the photon mode $(k,\lambda)$. To simplify our calculation we make the substitution (32) in (26) as well.

$$U(R) = \text{Re}\frac{i}{\pi}\left[\int_0^\infty 2(N(\omega)+1/2)\alpha_{eA}(\omega)\alpha_{gB}(\omega)\frac{\omega^4}{R^2}\left(1 + \frac{2i}{n(\omega)\omega R} - \frac{5}{(n(\omega)\omega R)^2}\right.\right.$$
$$\left. - \frac{6i}{(n(\omega)\omega R)^3} + \frac{3}{(n(\omega)\omega R)^4}\right)\exp(2in(\omega)\omega R)d\omega$$
$$+ \frac{2\pi i}{3}|d_{eg}^A|^2 2(N(\omega_A)+1/2)\alpha_{gB}(\omega_A)\frac{\omega_A^4}{R^2}$$
$$\left.\times\left(1 + \frac{1}{(n(\omega_A)\omega_A R)^2} + \frac{3}{(n(\omega_A)\omega_A R)^4}\right)\exp\left[-2\text{Im}(n(\omega_A))\omega_A R\right]\theta(\omega_A)\right].$$

This result is valid not only under thermal equilibrium, but for any electromagnetic field with the number of photons $N_{k\lambda}$ in the mode $(k,\lambda)$.

For a case of thermal equilibrium at temperature T we have $N(\omega) + \frac{1}{2} = \frac{1}{2}\coth\left(\frac{\omega}{2T}\right)$.

Consequently, if excited atom A interacts with ground-state atom B in a surrounding of thermal photons, we have

$$U(R) = \operatorname{Re} \frac{i}{\pi} \left[ \int_0^\infty \coth\left(\frac{\omega}{2T}\right) \alpha_{eA}(\omega) \alpha_{gB}(\omega) \frac{\omega^4}{R^2} \left(1 + \frac{2i}{n(\omega)\omega R} - \frac{5}{(n(\omega)\omega R)^2} \right. \right.$$

$$\left. \left. - \frac{6i}{(n(\omega)\omega R)^3} + \frac{3}{(n(\omega)\omega R)^4} \right) \exp(2i n(\omega)\omega R) d\omega \right.$$  (33)

$$\left. + \frac{2\pi i}{3} |d_{eg}^A|^2 \coth\left(\frac{\omega_A}{2T}\right) \alpha_{gB}(\omega_A) \frac{\omega_A^4}{R^2} \left(1 + \frac{1}{(n(\omega_A)\omega_A R)^2} + \frac{3}{(n(\omega_A)\omega_A R)^4} \right) \exp\left[-2\operatorname{Im}(n(\omega_A))\omega_A R\right] \theta(\omega_A) \right]$$

For the case of multi level atoms we arrive at

$$U(R) = \operatorname{Re} \frac{i}{\pi} \left[ \int_0^\infty \coth\left(\frac{\omega}{2T}\right) \alpha_A(\omega) \alpha_B(\omega) \frac{\omega^4}{R^2} \left(1 + \frac{2i}{n(\omega)\omega R} - \frac{5}{(n(\omega)\omega R)^2} \right. \right.$$

$$\left. \left. - \frac{6i}{(n(\omega)\omega R)^3} + \frac{3}{(n(\omega)\omega R)^4} \right) \exp(2i n(\omega)\omega R) d\omega \right.$$  (34)

$$+ \sum_m \frac{2\pi i}{3} |d_{0m}^A|^2 \coth\left(\frac{\omega_{0m}^A}{2T}\right) \alpha_B(\omega_{0m}^A) \frac{(\omega_{0m}^A)^4}{R^2}$$

$$\left. \times \left(1 + \frac{1}{(n(\omega_{0m}^A)\omega_{0m}^A R)^2} + \frac{3}{(n(\omega_{0m}^A)\omega_{0m}^A R)^4} \right) \exp\left[-2\operatorname{Im}(n(\omega_{0m}^A))\omega_{0m}^A R\right] \theta(\omega_{0m}^A) \right]$$

Where $\omega_{0m}^A = \varepsilon_0^A - \varepsilon_m^A$ is the transition frequency, 0 corresponds to the initial state of atom A, m corresponds to virtual state. The polarizabilities of atoms A (B) is

$$\alpha^{\nu\nu'}(\omega) = \sum_m \frac{d_{0m}^\nu d_{om}^{\nu'}}{\omega_{m0} - \omega - i\frac{\gamma_m}{2}} + \frac{d_{0m}^\nu d_{0m}^{\nu'}}{\omega_{m0} + \omega + i\frac{\gamma_m}{2}}.$$  (35)

The first term of (33) can be rewritten using contour integration [30]

$$U(R) = -\operatorname{Re} \left[ 2T \sum_{n=0}^\infty \alpha_{eA}(iu_n) \alpha_{gB}(iu_n) \frac{u_n^4}{R^2} \left(1 + \frac{2}{n(iu_n)u_n R} + \frac{5}{(n(iu_n)u_n R)^2} \right. \right.$$

$$\left. \left. + \frac{6}{(n(iu_n)u_n R)^3} + \frac{3}{(n(iu_n)u_n R)^4} \right) \exp(-2n(iu_n)u_n R)\left(1 - \frac{1}{2}\delta_{n0}\right) \right.$$  (36)

$$\left. + \frac{2}{3} |d_{eg}^A|^2 \coth\left(\frac{\omega_A}{2T}\right) \alpha_{gB}(\omega_A) \frac{\omega_A^4}{R^2} \left(1 + \frac{1}{(n(\omega_A)\omega_A R)^2} + \frac{3}{(n(\omega_A)\omega_A R)^4} \right) \exp\left[-2\operatorname{Im}(n(\omega_A))\omega_A R\right] \theta(\omega_A) \right]$$

Where $u_n = 2\pi n T$ is the Matsubara frequency.

If both atoms are in their ground-state, we should drop the second term. If the medium is absent ($n(\omega) = 1$), we find

$$U(R) = -\operatorname{Re} 2T \sum_{n=0}^\infty \alpha_{gA}(iu_n) \alpha_{eB}(iu_n) \frac{u_n^4}{R^2} \left(1 + \frac{2}{u_n R} + \frac{5}{(u_n R)^2} + \frac{6}{(u_n R)^3} + \frac{3}{(u_n R)^4} \right) \exp(-2u_n R)\left(1 - \frac{1}{2}\delta_{n0}\right)$$  (37)

So, we come to the result of papers [29-33].

### III. Interaction between two dilute gas media

We will consider interaction between two plates made of dissimilar dilute dielectrics if the amount of excited atoms is significant. Here, one should distinguish the two limiting cases [24]. The first case of short distance between the plates $LT \ll 1$, where L is the separation between the plates and T is temperature, has been considered in [17]. The violation of the Lifshitz formula for this case was demonstrated as a result of the presence of the excited atoms in the media. Here, we are going to consider another limiting case of large separations ($LT \gg 1$). To calculate the force, we take the interaction potential of two atoms (36) for excited atom and ground-state atom, and (37) for two ground-state atoms. If we took a potential, obtained in the framework of the perturbation theory, we would come to the divergence of integrals (29) for excited and ground-state atoms. But if we take the potential obtained with the help of non-perturbative method, the result is divergent no more. To simplify the calculations, we will use the following model. We suppose that only the atoms located no farther than the photon free mean path from the interface take part in the interaction. The photon free mean path is

$$L_{ph1} = \left(2\,\text{Im}\left(n(\omega_A)\right)\omega_A\right)^{-1},$$
$$L_{ph2} = \left(2\,\text{Im}\left(n(\omega_B)\right)\omega_B\right)^{-1} \quad (38)$$

So the integration is restricted by this distance. The exponent $\exp\left[-2\,\text{Im}\left(n(\omega_A)\right)\omega_A R\right]$ in the second term of the expression (36) will be dropped.

After calculation for $L \gg \lambda$ and

$$L \gg L_{ph} \quad (39)$$

the force per unit area is

$$F(L) = F_{res}(L) + F_{nres}(L). \quad (40)$$

The resonance force reads

$$F_{res}(L) = \frac{4\pi}{9} \frac{L_{ph1} L_{ph2}}{L} \frac{\left|d_{eg}^A\right|^2 \left|d_{eg}^B\right|^2 \omega_B \omega_A \left(\omega_B^2 - \omega_A^2\right)}{\left(\left(\omega_B^2 - \omega_A^2\right)^2 + \left(\gamma_B \omega_A\right)^2\right)} \left[\omega_A^3 n_A^e n_B^g \coth\left(\frac{\omega_A}{2T}\right) - \omega_B^3 n_A^g n_B^e \coth\left(\frac{\omega_B}{2T}\right)\right]. \quad (41)$$

To calculate the non-resonance force, we will consider the case of large distances and high temperatures ($LT \gg 1$).

$$F_{nres}(L) = \frac{\pi}{2L^3} T \alpha_{gA}(0) \alpha_{gB}(0) \left(n_A^g - n_A^e\right)\left(n_B^g - n_B^e\right), \quad (42)$$

Where $n_A^e\left(n_B^e\right)$, $n_A^g\left(n_B^g\right)$ are the density numbers of the excited atoms A (B) and ground state atoms A (B).

Let us compare the results obtained with the help of quantum electrodynamics (41) and (42) with the Lifshitz formula. For two plane media of dilute dielectrics the force, obtained using the Lifshitz formula [4,5] reads [34]

$$F_{Lif}(L) = \frac{\pi}{2L^3} T \alpha_{gA}(0) \alpha_{gB}(0) \left(n_A^g - n_A^e\right)\left(n_B^g - n_B^e\right)$$

This formula coincides with the non-resonance contribution (42).
If the density number of excited atoms is small, as it follows from (40), (41), and (42), the result of quantum electrodynamics coincides with the Lifshitz formula.

For a case of thermal equilibrium, the density numbers obey the Boltzmann distribution

$$n_A^e = n_A^g \exp\left(-\frac{\omega_A}{T}\right),\ n_B^e = n_B^g \exp\left(-\frac{\omega_B}{T}\right).$$

If we consider two-level atoms,

$$n_A^e + n_A^g = n_{0A},\ n_B^e + n_B^g = n_{0B},$$

where $n_0$ is the total density number of the corresponding medium.
Consequently

$$n_A^g = \frac{n_{0A}}{1+\exp(-\omega_A/T)},$$

$$n_A^e = \frac{n_{0A}\exp(-\omega_A/T)}{1+\exp(-\omega_A/T)}.$$

$$n_B^g = \frac{n_{0B}}{1+\exp(-\omega_B/T)},$$

$$n_B^e = \frac{n_{0A}\exp(-\omega_B/T)}{1+\exp(-\omega_B/T)}.$$

Finally, for the force of Casimir-Polder interaction, we have

$$F(L) = F_{Lif}(L) + \frac{4\pi}{9}\frac{L_{ph1}L_{ph2}}{L}\frac{\left|d_{eg}^A\right|^2 \left|d_{eg}^B\right|^2 \omega_B \omega_A \left(\omega_B^2 - \omega_A^2\right) n_{0A} n_{0B}}{\left[\left(\omega_B^2 - \omega_A^2\right)^2 + \left(\gamma_B \omega_A\right)^2\right]\left(1+\exp(-\omega_A/T)\right)\left(1+\exp(-\omega_B/T)\right)} \quad (43)$$

$$\times \left[\omega_A^3 \exp(-\omega_A/T)\coth\left(\frac{\omega_A}{2T}\right) - \omega_B^3 \exp(-\omega_B/T)\coth\left(\frac{\omega_B}{2T}\right)\right]$$

$$F_{Lif}(L) = \frac{2\pi}{9L^3}\frac{T\left|d_{ge}^B\right|^2 \left|d_{ge}^A\right|^2 n_{0A} n_{0B}}{\omega_A \omega_B}\tanh\left(\frac{\omega_A}{2T}\right)\tanh\left(\frac{\omega_B}{2T}\right) \quad (44)$$

The photon free mean path of the dilute media can be calculated using equations (38), (22), (19) and $\varepsilon(\omega) = 1 + 4\pi n_0 \alpha_g(\omega)$.

$$L_{ph1} = 3\frac{\left(\omega_B^2 - \omega_A^2\right)^2 + \left(\gamma_B \omega_A\right)^2}{4\pi n_B^g \left|d_{ge}^B\right|^2 \gamma_B \omega_A^2}, \quad (45)$$

$$L_{ph2} = 3\frac{\left(\omega_B^2 - \omega_A^2\right)^2 + \left(\gamma_A \omega_B\right)^2}{4\pi n_A^g \left|d_{ge}^A\right|^2 \gamma_A \omega_B^2}. \quad (46)$$

Where $\gamma_A$ and $\gamma_B$ are collisional widths of the excited states of atoms A and B, which can be calculated as follows [35]

$$\gamma_A = \gamma_{ANat} + n_{0A}k_{Abr},$$
$$\gamma_B = \gamma_{BNat} + n_{B0}k_{Bbr}.$$

Here $\gamma_{Nat}$ is the natural width of excited level, $k_{br}$ is the broadening rate coefficient [35], which does not depend on the density number of the medium.
Substituting (45) and (46) into (43), we find

$$F(L) = F_{Lif}(L) + \frac{\left(\omega_B^2 - \omega_A^2\right)^3 \left[\omega_A^3 \exp(-\omega_A/T)\coth\left(\frac{\omega_A}{2T}\right) - \omega_B^3 \exp(-\omega_B/T)\coth\left(\frac{\omega_B}{2T}\right)\right]}{4\pi L \omega_A \omega_B n_{0A} n_{0B} k_{Abr} k_{Bbr}\left(1+\exp(-\omega_A/T)\right)\left(1+\exp(-\omega_B/T)\right)} \quad (47)$$

Here, we neglected the natural widths of excited levels of atoms. For a case of low density numbers the collisional widths can be neglected in comparison with the natural one, and the second term of (47) does not depend on the density numbers. Consequently, if we put the density numbers equal to zero, the second term of (47) seams to survive. But if $n_0 = 0$ the inequality (39) is violated, thus the formula (47) can not be applied for the case of small density numbers.
Let $T \ll \omega_A, \omega_B$, then the number of excited atoms is low and the second term of (47) is negligible, thus, expression (47) coincides with the Lifshitz formula. It means that the force is attractive. If we consider a case of the media of like atoms, namely, $\omega_A = \omega_B$, we obtain the Lifshitz formula as well. Strictly speaking we can not implement the above develop method for identical atoms, for one should take into account resonance coupling between the atoms which leads to new symmetric or antisymmetric eigenstates of the pair of atoms [36], but after averaging over all possible orientations of the dipole moments of the atoms this effect will be canceled. Consequently, for identical media we come to the Lifshitz formula. The difference between expression (47) and the Lifshitz formula is significant for low density numbers or for dilute gas media for the density numbers are in the denominators of the second term of (47). We expect that the difference between the Lifshitz formula and the result of QED will disappear if the density numbers are large, so that the media could be considered as continuous ones. Let $T \gg \omega_A, \omega_B$, then the expression (47) can be rewritten as follows

$$F(L) = F_{Lif}(L) - \frac{T\left(\omega_B^2 - \omega_A^2\right)^4}{8\pi L \omega_A \omega_B n_{0A} n_{0B} k_{Abr} k_{Bbr}} \tag{48}$$

If the second term is dominant, the interaction between the media is repulsive.

### IV. Summary

We have shown that the perturbation technique can not be applied to the problem of Casimir interaction between two media containing excited atoms even if the media are diluted. The reason for this inapplicability is as follows. Obtained in the framework of the perturbation method, the interaction potential between a pair of atoms (if one of them is excited) for the retarded regime drops as $R^{-2}$ with the distance between the atoms [25,26]. If we took into account only pair interactions between the atoms of the media, we would have to sum the interaction potentials over all the pairs of atoms. This summation results in divergence.

To handle the divergence, we developed a non-perturbative approach to the problem. We considered interaction between two atoms embedded in an absorbing dielectric medium. The result we obtained differs from the perturbation one in two respects. First, we took into account the finite widths of excited levels of the atoms. Second, we included possible absorption of real photons which are responsible for the interaction between an excited atom and a ground-state atom in retardation regime. It results in the exponential factor $\exp\left[-2\operatorname{Im}(n(\omega_A))\omega_A R\right]$ depending on the imaginary part of the refractive index of the medium. Now, if we sum up all the pair potentials for two media, we will come to divergence no more. For two ground-state atoms embedded in a dielectric medium we obtained the results of [27]. We calculated the Casimir force between two planes of dilute gas media. The temperatures are high enough for the media to contain excited atoms. We considered a retarded regime for the non-retarded one was treated before [17]. For $LT \gg 1, L \gg \lambda$, and $L \gg L_{ph}$ we obtained expression (47). We had to introduce a new parameter – the free mean path of the photon in the medium $L_{ph}$ – to this problem, which characterizes possible absorption of the photon in the interacting media.

The result of quantum electrodynamics (47) differs from the Lifshitz formula which was intended for a case of continuous media [24] ($n_0 \lambda^3 \gg 1$). Here, we considered an opposite case of dilute gas media ($n_0 \lambda^3 \ll 1$). Although, the results of the Lifshitz formula and quantum electrodynamics coincide for two dilute gas media at low temperatures (if the number of excited atoms is negligible) [24], the coincidence is violated for the case of high temperatures (if the number of exited atoms if significant). The similar result was obtained in [17] for non-retarded case.

The difference between the Lifshitz formula and the result of quantum electrodynamics is in the second term of (47). If both media are composed of the identical atoms $\omega_A = \omega_B$ the second term of (47) is equal to zero and the Lifshitz formula coincided with the QED approach. The density numbers of the media are in the denominator of the second term, while they are in the numerator of the Lifshitz formula(44). Thus, if the density numbers are high, the difference between the formulas disappears. So for dense media $n_0 \lambda^3 \gg 1$ both approaches lead to the same results.

For a case of high temperatures $T \gg \omega_A, \omega_B$, we derived formula (48). If the density numbers are small, the force becomes repulsive, while the Lifshitz formula leads to attraction.

### References

1. F. London, 1930, *Z.Phys.* **63**, 245.
2. H. B.G. Casimir and D. Polder, 1948, *Phys. Rev.* **73**, 360.
3. H. B.G. Casimir, 1948 *Proc.K.Ned.Akad.Wet.Ser.B* **51**, 793.
4. E. M. Lifshitz, 1956 *Sov. Phys. JETP* **2**, 73.
5. I.E. Dzyaloshinskii , E.M. Lifshitz and L.P. Pitaevskii, 1960 *Sov.Phys. JETP* 10, 161.
6. K. Milton, 2004 *J. Phys. A* **37**, R209; M. Bordag, U. Mohideen and V. M. Mostepanenko, 2001, *Phys. Rep* **353**, 1; S.Buhmann, D.-G. Welsch. 2006 *quant-ph/0608118*
7. D.L.Nelson, m.M. Cox, 2002 *Lehninger principles of biochemistry* (Worth, New York), pp. 177, 250-253, 331, 365, 1053.
8. K. Autumn, M. Sitti, Y. A. Liang, A.M. Peattie, W. R. Hansen, S. Sponberg, T. W. Kelly, R. Fearing, J.N. Israelachvili, R. J. Full, 2002 *Proc. Natl. Acad. Sci. USA* **99** (19), 12252.
9. K. B. Kesel, A. Martin, T. Seidl, 2004 *Smart Mater. Struct*. **13**, 512.
10. Th. Foster 1965 *in Modern Quantum Chemistry*, ed O. Sinanoglu (Academic, New York), Pt.3.
11. H. van Amerogen, L Valkunas. Grondelle, 2000 *Photosynthetic Excitons*,(World Scientific).
12. E. Haas, M. Wilchek, E. Katchalski-Katzir, and I. Z. Steinberg, 1975, *Distribution of end-to-end distances of oligopeptides in solution as estimated by energy transfer*:*Proc. Nat. Acad. Sci. USA* **72**, 1807.
13. L. Stryer, Fluorescence energy transfer as a spectroscopic ruler, 1978 *Ann. Rev. Biochem.* **47**, 819.
14. S. Weiss, Fluorescence spectroscopy of single biomolecules, 1999 *Science* **283**,1676.
15. D. M. Leitner, Vibrational energy transfer in helices, 2001 *Phys. Rev. Lett.* **87**,188102.
16. A. E.Cohen, 2003 *Ph.D thesis* (Cambridge, UK).
17. Yury Sherkunov, 2005 *Phys. Rev. A* **72**, 052703.
18. M. Fichet, F. Shuller, D. Bloch, and M. Ducloy, 1995, *Phys. Rev. A* **51**, 1553; H. Failache, S. Saltiel, M. Fitchet, D. Bloch, and M. Ducloy, 1999, *Phys. Rev. Lett.* **83**, 5467;H. Failache, S. Saltiel, A. Fischer, D. Bloch, and M. Ducloy, 2002, *Phys. Rev. Lett.* **88**, 24603-1.
19. J. M. Wylie and J. E. Sipe, Phys. Rev. A **30**, 1185 (1984); J. M. Wylie and J. E. Sipe, Phys. Rev. A **32**, 2030 (1985); S. Y. Buhmann, L. Knoll, D.- G. Welsch, and Ho Trung Dung, Phys. Rev. A **70**, 052117 (2004).
20. L. V. Keldysh, 1964 *Sov. Phys. JETP* **20**, 1018.
21. E. M. Lifshitz and L. P. Pitaevski 1981, *Physical kinetics* (Pergamon, Oxford).
22. B. A. Veklenko, 1989 *Sov. Phys. JETP* **69**, 258.
23. V.B. Berestetski, E.M. Lifshitz, and L.P. Pitaevski 1982, *Quantum electrodynamics* (Pergamon, Oxford).
24. E. M. Lifshitz and L.P.Pitaevski, 1980, *Statistical physics. Part 2* (Butterworth-Heinenmann, Oxford).
25. E. A. Power and T. Thirunamachandran, 1993 *Phys. Rev. A* **47**, 2539.
26. E. A. Power and T. Thirunamachandran, 1995 *Phys. Rev. A* **51**, 3660.
27. S. Buhmann, H. Safari, and D.-G. Welsch, 2006 *Preprint quant-ph*/0603193; H. Safari, S. Y. Buhmann, D.-G. Welsch, and H. T. Dung 2006 *Phys. Rev. A* **74**, 042101; M. S. Tomas, 2005, *Phys. Rev. A* **72**, 034104 .
28. P.W. Milonni and M.-L. Shih, 1992 *Phys.Rev.A,* **45**, 4241.
29. B.W. Ninham and J. Daicic, 1999 *Phys. Rev. A*, **57**, 1870.
30. G.H. Goedecke and Roy C. Wood, 1999 *Phys.Rev A*, **60,** 2677.
31. H. Wennerstrom, J. Daicic, and B.W. Ninham, 1999 *Phys. Rev. A*, **60**, 2581.
32. G. Barton, 2001 *Phys. Rev. A*, **64**, 032102.
33. Larry Spruch, 2002 *Phys. Rev. A*, **66**, 022103.
34. B. Geyer, G.L. Klimchitskaya, and V.M. Mostepanenko 2005 *Phys. Rev. D*, **72**, 085009.
35. C.G. Carrington, D.N. Stacey, and J. Cooper, 1973 *J. Phys. B*, **6**, 417.
36. M. Bostrom and B. W. Ninham, 2004 *Phys. Rev. A* **69**, 054701.